\documentclass[prb,aps,twocolumn,showpacs,superscriptaddress,amsmath,amssymb,floatfix]{revtex4}
\usepackage{graphicx}

\newcommand{\be}{\begin{equation}}
\newcommand{\ee}{\end{equation}}

\newcommand{\eq}[1]{(\ref{#1})}

\begin{document}
\title{The stability of nonequilibrium polariton superflow in the presence of a cylindrical defect}
\author{Michiel Wouters}
\affiliation{TQC, Universiteit Antwerpen, Universiteitsplein 1, 2610 Antwerpen, Belgium}
%\author{Vincenzo Savona}
%\affiliation{Insitute of Theoretical Physics, Ecole Polytechnique F\'ed\'erale de Lausanne (EPFL), CH-1015 Lausanne, Switzerland}
\begin{abstract}
We make a theoretical study of the stability of nonequilibrium polariton superflows that interact with a cylindrical defect. The nonresonantly pumped polariton condensate is modelled with a generalized complex Ginzburg-Landau equation. At low pump intensities the dissipation is found stabilize the superflow. At large pump intensities, we find an instability that sets a lower critical speed for superfluidity. For even larger pump power, the lower and upper critical speed meet and stable superflows are no longer possible.
\end{abstract}
\maketitle

\section{Introduction}

The superfluid properties of exciton polariton quantum fluids is a topic of active experimental and theoretical investigation. As predicted by Carusotto an Ciuti~\cite{iac-superfl}, recent experiments by Amo et al.~\cite{amo-superfl} have shown that the scattering of the polaritons off a weak defect is strongly suppressed when the polariton speed is below the Landau critical speed. More recently, the scattering with a large defect was addressed theoretically~\cite{simon_vortex} and experimentally under continuous wave~\cite{amo_vortex} and pulsed~\cite{gael_vortex} excitation. In contrast to the case of weak defects, where the supercurrents emit sound waves, strong defects lead to the nucleation of vortices and solitons.

An important aspect of polariton condensates is the finite polariton life time, that makes a profound difference with equilibrium Bose-Einstein condensates of dilute vapors of ultracold atomic gases. When the finite polariton life time is compensated by continuous resonant excitation, the driving laser fixes the phase of the polariton fluid, which prevents the formation of phase defects such as vortices. In order to circumvent this phase rigidity, the defect was placed behind the excitation spot~\cite{amo_vortex}. An alternative possibility to replenish the polariton condensate without fixing the phase of the condensate is by so-called nonresonant excitation. High energy excitations are injected by the laser and during their relaxation toward the bottom of the lower polariton branch, they loose all phase information and spontaneously form a coherent state when the density exceeds a threshold~\cite{kasprzak}.

Where the superfluid properties under resonant excitation are remarkably close to the ones of equilibrium condensates, the nonresonant driving introduces profound differences. Because the decay rate is not negligible with respect to the interaction energy, the driven-dissipative nature of the polariton gas strongly affects the excitation spectrum~\cite{szy,nonresonant} of polariton condensates and hence their superfluid properties. In a previous work, we have studied the faith of polariton superflows when they scatter with weak defects~\cite{superfluid-noneq}. We have found that the dissipation actually stabilizes the superflow and that the critical velocity is higher than in an equilibrium condensate at the same density. Even when excitations are created and a drag force is exerted on the defect, the condensate is not destroyed, but is able to maintain its superfluid velocity. A finite concentration of defects is needed to destroy the superflow. 

From the theoretical side, much of the behavior of equilibrium superfluids can be understood from the Gross-Pitaevskii equation (GPE) that describes the order parameter of a quantum degenerate Bose gas~\cite{SSLP-book}. In particular, the GPE provides a description of the onset of a drag force on a defect that moves in a superfluid that sets on only when the defect defect velocity exceeds a critical speed $v_c$. For a weak defect, the GPE description reproduces the Landau criterion $v_c=c$, where $c$ is the speed of sound~\cite{grisha}. Above the critical speed, the emission of phonons is responisible for a drag force on the defect.

The Gross-Pitaevskii equation is not limited to the limit of weak defects and allows for the investigation of nonperturbative defects that are dragged through a superfluid. Carrying out this analysis, Frisch et al. found that the critical velocity of a large cylinder is approximately two times smaller than the speed of sound~\cite{french}. In that case, the drag force on the defect results from the emission of vortex anti-vortex pairs~\cite{winiecki}, that were recently observed in a Bose Einstein condensate of ultracold atoms~\cite{neely}. Above the critical speed, the continuous generation of excitations by the defect heats up the Bose gas, resulting in a final state of the gas moving along with the defect.

The Gross-Pitaevskii equation can be straightforwardly generalized to nonequilibrium polariton condensates, where the steady state is determined by a balance between driving and dissipation. Supplementing it with dissipative terms leads to the so-called complex Ginzburg-Landau equation (cGLE)~\cite{aranson}, first used in the context of polariton condensation in Ref.~\onlinecite{keeling}.

In this work, we use the cGLE to analyze the collision of a polariton condensate with a large diameter high cylindrical potential. For small dissipation, we find that the equilibrium picture continues to hold qualitatively: the superflow decays above a certain critical velocity. In analogy with the case of a perturbative defect, the critical velocity is increased by the dissipation. Surprisingly, we will find at that large dissipation, superfluidity is completely destroyed by a strong defect. The reason is a nontrivial pattern of the currents in the condensate. In an intermediate window of dissipation, superflows are stable between a lower and upper critical velocity.

The outline of the paper is as follows. In Sec.~\ref{sec:model}, we recapitulate the generalized complex Ginzburg Landau Equation (cGLE) that we use for the analysis of nonequilibrium superfluidity. In Sec.~\ref{sec:cond}, we describe the condensate state in the presence of a large cylindrical defect. The stability of supeflows is then investigated in Sec.~\ref{sec:sup}. Conclusions are finally drawn in Sec.~\ref{sec:concl}.

\section{Theoretical model \label{sec:model}}

We use in this work a generalization of the cGLE, that we have introduced in Ref.~\onlinecite{superfluid-noneq}. It includes a frequency-dependence of the pumping:
\begin{multline}
	i \frac{d\psi}{dt} = \left\{ -\frac{\hbar}{2m}\nabla^2 + V_{ext} \right. \\
			\left.	+ \frac{i}{2}\left[ P\,\left(1- \frac{i}{\Omega_K}  \frac{d}{dt}\right)-r |\psi|^2
			-\gamma\right]
			        +  g |\psi|^2  \right\} \psi.
	\label{eq:CGLE}
\end{multline}
The energy zero has been set for convenience at the bottom of the lower polariton branch; the efficiency of amplification (proportional to the pumping strength $P$) decreases to zero in a frequency interval $\Omega_K$ above it. Assuming a linear form of the frequency dependence of the amplification, the generalized GPE \eq{eq:CGLE} maintains a temporally local form.
The others terms describe gain saturation ($r$), losses ($\gamma$), polariton mass ($m$), polariton-polariton interactions ($g$),  external potential  ($V_{ext}$).

If the strength of pumping is large enough to overcome losses $P>\gamma$, the $\psi=0$ state is dynamically unstable against the creation of a finite condensate amplitude in any of the the low-momentum modes for which $\hbar k^2/2m <\Omega_K\,(1-\gamma/P)$. The rate of this instability is maximum at $k=0$ and decreases for increasing $k$. Under these conditions, a condensate is formed at $k=0$.

\section{condensate state \label{sec:cond}}

In the presence of an external potential $V_{ext}$, the condensation phenomenology becomes much richer. If $V_{ext}$ is a random potential, that naturally occurs because of the fluctuations during the microcavity growth, the formation of multiple condensates was observed and explained within a related theoretical formalism~\cite{disorderPRB}.

In this work, we will study the effect of a cylindrical defect on the condensate. The amplitude is taken to be much larger than the polariton blue shift $g |\psi|^2$ due to the polariton-polariton interactions. In the absence of driving and dissipation, this type of potential makes a hole in the condensate that recovers to the asymptotic density on a distance fixed by the healing length $\xi=\sqrt{\hbar^2/m g n_c}$, where $n_c$ is the condensate density.

\begin{figure}[htbp]
\begin{center}
\includegraphics[width=\columnwidth,angle=0,clip]{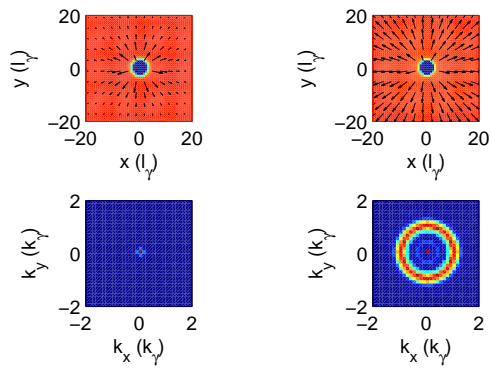}  % made with figure4.m
\end{center}
\caption{Real (upper panels) and momentum (lower panels) space of a spontaneously formed polariton condensate in the presence of a defect. {\em Left hand panels --} Weakly out of equilibrium (small pumping and gain saturation), the condensate forms at $k=0$. {\em Right hand panels --} far from equilibrium, the condensate forms on a ring in momentum space, because of the currents away from the defect. Paramters: $P/\gamma=2$ (left) and $P/\gamma=15$ (right); $r/g=3$, defect with diameter of $5 l_\gamma$ and hight of $10\gamma$. Lengths and momenta are expressed in terms of $l_\gamma=\sqrt{\hbar^2/(m\gamma)}$ and $k_\gamma=l_\gamma^{-1}$ respectively.}
\label{fig:instab_k0}
\end{figure}

When dissipation is included, this simple state can be dramatically modified. Steady states of the condensate state for two values of the pump power are shown in Fig.~\ref{fig:instab_k0}. It is striking for the larger pump intensity, the condensate does no longer form close to zero momentum, but instead on a circle in momentum space. The arrows in the real space picture show the flow of the condensate and indicate that the condensate flows away from the defect.  The physical explanation for the condensate flow is the following. The strong defect forces the density to vanish. Close to the defect, the density is below the value where the gain is saturated. Stabilization can then only occur when the excess created particles are carried away by a particle flow. This flow forces the condensation to occur at finite momentum even far away from the defect. This interpretation supports the numerical observation that the radius of the ring in momentum space increases with increasing pump power.

The finite momentum states in Fig.~\ref{fig:instab_k0}  present flows of the condensate polaritons that are the gradient of the superfluid phase $v_s=\nabla \theta$ and one may be tempted to conclude that the observation of such flows would prove superfluidity. The analogy of this type of superflows with superfluidity phenomena in equilibrium Bose gases is however difficult to sustain and strongly relies on the fact that energy is fed into the system. It is therefore analogous to the flow of water with a pressure difference between both ends.

\section{Stability of superflow \label{sec:sup}}

\begin{figure}[htbp]
\begin{center}
\includegraphics[width=\columnwidth,angle=0,clip]{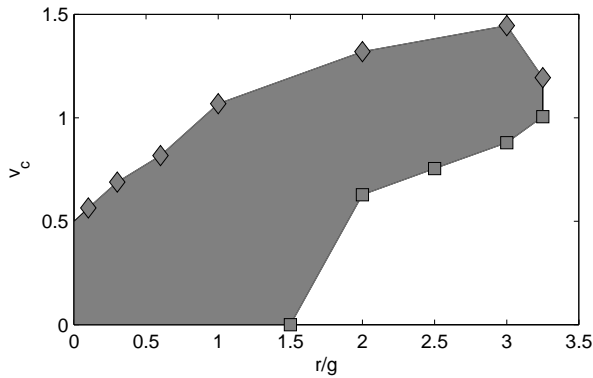}  % made with figure4.m
\end{center}
\caption{Critical velocity of a nonequilibrium condensate flowing against a big defect as a function of the gain saturation parameter $r$. Diamonds denote the standard critical velocity for breaking due to vortex emission, where the squares denote the lower critical velocity, below which the condensate state is unstable with respect to the formation of a ring in momentum space (see Fig. \ref{fig:lower_vc}). Parameters: $g |\psi|^2=1.5 \gamma$, same defect as in Fig.~\ref{fig:instab_k0} }
\label{fig:vcsd}
\end{figure}

A true manifestation of superfluidity is the possibility of manifesting multiple flow patterns. In the absence of defects, flow of equilibrium superfluids is stable at any velocity, thanks to Galilean invariance. When a weak defect potential breaks this symmetry, the superflow loses stability above the speed of sound $v_c=\sqrt{g n_c/m}$. For a large cylindrical defect, the critical velocity is smaller and equals~\cite{french} $v_c = \sqrt{2/11} c$.

In polariton condensates, condensation can be forced to occur in a momentum state different from zero by using a strong seed pulse at a finite wave vector. This acts as an initial condition for the condensate field and the stability of the superflow will determine whether the condensate continues to flow at that speed or will rather relax to a lower momentum. For simplicity, we work with periodic boundary conditions so that the flowing condensate does not leave the region where the cylinder is present. In practice this geometry could be realized with a ring shaped trapping potential.

In our previous work~\cite{superfluid-noneq}, we have investigated the stability of superflow for driven-dissipative condensates. We have pointed out that the energy dependence of the gain term breaks Galilean invariance and that therefore the critical velocity is finite even in the clean system. The effect of defects is on the other hand weakened by the dissipation. The critical velocity may be well above the one predicted for the equilibrium condensates $v_c=\sqrt{g n_c/m}$.

We will now turn to the investigation of the critical velocity when the nonequilibrium superfluid collides with a large cylinder. We have found that a single strong defect can lead to the nucleation of vortices and hence to the decay of the superflow. In order to assess the effect of the nonequilibrium situation on the stability of the superflow, we plot with diamonds in Fig.\ref{fig:vcsd} the critical velocity as a function of the parameter $r/g$, keeping the condensate density constant. 
Let us first concentrate on small values of the gain saturation parameter $r/g$.
For $r/g$ tending to zero, we recover the equilibrium critical velocity $v_c\approx0.5 c$. The critical velocity increases with the dissipation parameter $r/g$. In analogy to the case of weak defects, the dissipation stabilizes the superflow. 

\begin{figure}[htbp]
\begin{center}
\includegraphics[width=\columnwidth,angle=0,clip]{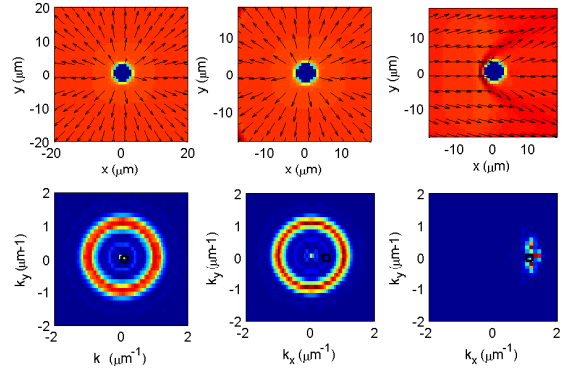} % made with figure3.m
\end{center}
\caption{Time evolved nonequilibrium condensate colliding with a big defect for various initial wave vectors, with increasing initial velocity from left to right. The upper panels show the real space density and flow (arrows), the lower panels show the momentum distribution. The squares indicate the initial wave vector of the condensate. Only when the initial velocity overcomes a critical value, the condensate maintains its momentum. For low initial velocity, a flow away from the defect appears.}
\label{fig:lower_vc}
\end{figure}

When the gain saturation is increased, the pump power should also increase to keep the density constant. At a certain value of the gain saturation ($r/g\approx 1.5$ for the density considered here), the condensate starts to form on a ring in momentum space with radius $k_c$, as discussed above. From our numerical simulations, we have found that superflows are only stable if their wave vector is larger than $k_c$. We thus find a lower critical velocity for the superflow, that is indicated with squares in Fig.~\ref{fig:vcsd}. The region of stable superflow is the shaded area. 

For large values of $r/g$, the upper critical velocity decreases with gain saturation. At a critical value of $r/g$ ($\approx 3.5$ for the considered density), the upper and lower critical velocity coincide and a stable superflow can no longer be generated. 
We assume that the physical reason is the following. The speed in the wake behind the defect is larger than the asymptotic speed that is imposed by the initial condition. For increasing gain saturation (and pumping so to keep the density fixed), the local velocity increases and therefore exceeds the critical velocity for lower values of the asymptotic velocity.

\begin{figure}[htbp]
\begin{center}
\includegraphics[width=\columnwidth,angle=0,clip]{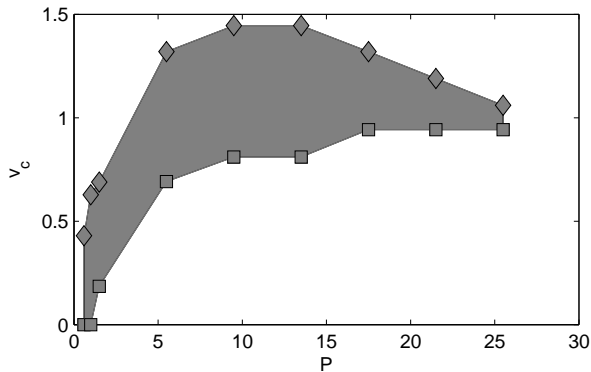} % made with figure3.m
\end{center}
\caption{Lower (squares) and upper (diamonds) critical velocity of a nonequilibrium condensate as a function of the pumping $P$. The gain saturation $r=0.3$ and interaction strength $g=0.1$.}
\label{fig:P_vc}
\end{figure}

This analysis has highlighted the role of gain saturation on the superfluid behavior of nonequilibrium condensates. In experiments however, it is rather the pump intensity than the gain saturation that is varied. In Fig.~\ref{fig:P_vc}, we show the lower and upper critical velocity as a function of the pump power, for fixed $r$ and $g$. It shows qualitatively the same trend as Fig.~\ref{fig:vcsd}. The increase of the upper critical velocity with pump power is now coming from both the increased role of interaction energy $g|\psi|^2$ and gain saturation $r|\psi|^2$. The decrease of the upper critical velocity at high powers is due to the effect of dissipation described above. The lower critical velocity is zero for small powers ($P\leq 1$), increases and then saturates until it meets the upper critial speed.

\begin{figure}[htbp]
\begin{center}
\includegraphics[width=\columnwidth,angle=0,clip]{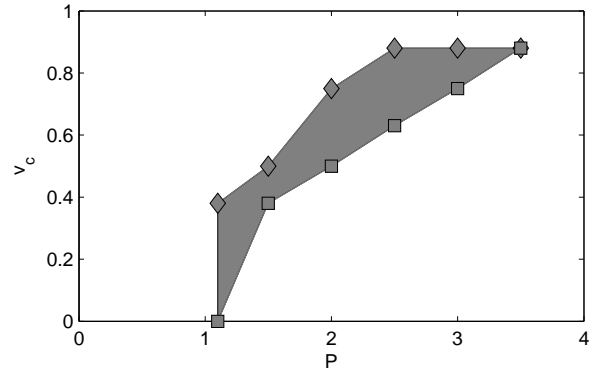} % made with figure5.m
\end{center}
\caption{The same as Fig.~\ref{fig:P_vc}, but with zero interaction strength $g=0$.}
\label{fig:noint}
\end{figure}

The enhanced stability of the superflow in nonequilibrium condensates with respect to the equilibrium ones raises the question whether the nonequilibrium superflow can be stable even without interactions between the polaritons ($g=0$). Fig.~\ref{fig:noint} shows that this is indeed the case. A stability region between the lower and upper critical velocity exists for pump intensities not too much above threshold. Gain saturation alone thus provides a nonlinearity that is sufficient to make superflows stable agains scattering off defects. This observation is in stark contrast with equilibrium condensates where the only nonlinearity comes from interactions and therefore the noninteracting Bose gas has a zero critical velocity.

\section{Conclusions \label{sec:concl}}

We have investigated the faith of superflows of nonequilibrium polariton condensates when they collide with a cylindrical defect.
The combination of the driven dissipative nature of polariton condensates with the defect potential can have a dramatic effect on the condensate shape. Above a critical pump power, the condensate is formed on a ring in momentum space, rather than around $k=0$. 
When a superflow is imposed on the condensate, a rich phenomenology is found. For small pump power/gain saturation, the equilibrium phenomelogy is recovered: stable superflow below a critical velocity  and shedding of vortices above it. For increasing driving, the $k=0$ state becomes unstable and superflows are only stable above a lower critical velocity. At large driving, the upper critical velocity decreases and finally meets the lower one at the critical power above which no longer a stable superflow exists. Finally, we have shown that even in the absence of polariton-polariton interactions, a finite critical speed exists thanks to the gain saturation nonlinearity.

\section{Acknowledgments}
I am indebted to I. Carusotto, who encouraged me to work on this problem. Stimulating discussions with V. Savona are acknowledged. This work was financed by the Fund for Scientific Research Flanders (FWO) projects G.036508, G.0370.09 and G.0180.09.


\begin{thebibliography}{99}

\bibitem{iac-superfl} I. Carusotto and C. Ciuti, Phys. Rev. Lett. {\bf 93}, 166401 (2004).

\bibitem{amo-superfl}  A. Amo {\em et al.}, Nat. Phys. {\bf 5}, 805 (2009).

\bibitem{simon_vortex} S. Pigeon, I. Carusotto and C. Ciuti, Phys. Rev. B {\bf 83}, 144513 (2011).

\bibitem{amo_vortex} A. Amo {\em et al.}  arXiv:1101.2530.

\bibitem{gael_vortex} G. Nardin,	{\em et al.}, Nat. Phys. advanced online publication  doi:10.1038/nphys1959 (2011).

\bibitem{kasprzak} J. Kasprzak {\em et al.}, Nature {\bf 443}, 409 (2006).


\bibitem{szy} M. H. Szyma\'{n}ska, J. Keeling, and P. B. Littlewood,
Phys. Rev. Lett. \textbf{96}, 230602 (2006).

\bibitem{nonresonant} M. Wouters and I. Carusotto, Phys. Rev. Lett. {\bf 99}, 140402 (2007).

\bibitem{superfluid-noneq} M. Wouters and I. Carusotto, {\bf 105}, 020602 (2010).


\bibitem{SSLP-book} L.P. Pitaevskii and S. Stringari, \textsl{Bose-Einstein
Condensation}, Clarendon Press Oxford (2003).

\bibitem{grisha}  G. E. Astrakharchik and L. P. Pitaevskii, Phys. Rev. A {\bf 70}, 013608 (2004).

\bibitem{french} T. Frisch, Y. Pomeau and S. Rica, Phys. Rev. Lett. {\bf 69}, 1644 (1992).

\bibitem{winiecki} T. Winiecki, J. F. McCann, and C. S. Adams, Phys. Rev. Lett. {\bf 82}, 5186 (1999).

\bibitem{neely} T. W. Neely, E. C. Samson, A. S. Bradley, M. J. Davis, and B. P. Anderson, Phys. Rev. Lett. {\bf 104}, 160401 (2010).

\bibitem{aranson} I. S. Aranson and Lorenz Kramer, Rev. Mod. Phys. {\bf 74}, 99 (2002).

\bibitem{keeling} J. Keeling and N. G. Berloff, Phys. Rev. Lett. {\bf 100}, 250401 (2008).

\bibitem{disorderPRB} D. N. Krizhanovskii, K. G. Lagoudakis, M. Wouters, B. Pietka, R. A. Bradley, K. Guda, D. M. Whittaker, M. S. Skolnick, B. Deveaud-Plédran, M. Richard, R. André, and Le Si , Phys. Rev. B {\bf 80}, 045317 (2009).


\end{thebibliography}
\end{document}